\documentclass{jfm}

\usepackage{graphicx}
\usepackage{natbib}

\newcommand{\RomanNumeralCaps}[1]
\linenumbers

\usepackage{bm,amsmath,amsfonts,color,amssymb,dcolumn,siunitx,umoline,caption,adjustbox,pdflscape,setspace}
\usepackage[utf8]{inputenc}
\usepackage{wasysym}
\usepackage{verbatim}
\usepackage[normalem]{ulem}

\title{Wall-bounded periodic snap-through and contact of a buckled sheet}
\author{Ehsan Mahravan\aff{1}, Mohsen Lahooti\aff{2} and Daegyoum Kim\aff{1}\corresp{\email{daegyoum@kaist.ac.kr}}}
\shorttitle{Wall-bounded periodic snap-through and contact}
\shortauthor{E. Mahravan, M. Lahooti and D. Kim}

\affiliation{\aff{1}Department of Mechanical Engineering, KAIST, Daejeon 34141, Republic of Korea
\aff{2} School of Engineering, Newcastle University, Newcastle upon Tyne, NE1 7RU, United Kingdom}

\begin{document}

\maketitle

\begin{abstract}
Fluid flow passing a post-buckled sheet placed between two close confining walls induces periodic snap-through oscillations and contacts that can be employed for triboelectric energy harvesting. The responses of a two-dimensional sheet to a uniform flow and wall confinement in both equilibrium and post-equilibrium states are numerically investigated by varying the distance between the two ends of the sheet, gap distance between the confining walls, and flow velocity. Cases with strong interactions between the sheet and walls are of most interest for examining how contact with the walls affects the dynamics of the sheet and flow structure. At equilibrium, contact with the wall displaces the sheet to form a nadir on its front part, yielding a lower critical flow velocity for the transition to snap-through oscillations. However, reducing the gap distance between the walls below a certain threshold distinctly shifts the shape of the sheet, alters pressure distribution, and eventually leads to a notable delay in the instability. The contact between the oscillating sheet and the walls at post-equilibrium is divided into several distinct modes, changing from sliding/rolling contact to bouncing contact with increasing flow velocity. During this transition, the time-averaged contact force exerted on the sheet decreases with the flow velocity. The vortices generated at the extrema of the oscillating sheet are annihilated by direct contact with the walls and merging with the shear layers formed by the walls, resulting in a wake structure dominated by the unstable shear layers.
\end{abstract}

\section{Introduction}
\label{sec:introduction}

High-amplitude vibrations of an elastic sheet in a fluid flow provide a novel mechanical approach for harvesting fluid kinetic energy. Diverse configurations of a flapping flag with one fixed end and one free end have been suggested for applications to piezoelectric energy harvesters. These are mostly based on the deformation of single or multiple flags \citep{Doar2011,Tosi2021}, a flag with an upstream bluff body \citep{ALLEN2001,Akaydin2010,Mittal_2022}, single or multiple inverted flags \citep{Kim2013,Shoele2016,Ryu2018,Tavallaeinejad2020,Mazharmanesh2020}, and an inverted flag with an upstream bluff body \citep{Hyeonseong_2017}. Furthermore, for triboelectric energy generation, the contact and separation of a flag with a nearby rigid wall \citep{Bae2014,Meng2014,Zhang2020,Lee_2021} or bluff body \citep{Zhang2020} and the mutual contact and separation of two side-by-side flags \citep{Chen2020} have been investigated.

A buckled elastic sheet with two clamped ends, which is initially bi-stable, rapidly snaps to the other side if the external force exerted on the sheet satisfies a certain criterion. This snap-through motion can be initiated by various external inputs, including a point force \citep{Chen2011,Pandey2014}, a change in the supporting angle of the sheet \citep{Beharic2014, Gomez2017_1}, the capillary force of a droplet deposited on the sheet \citep{Fargette2014}, and a midpoint magnetic force \citep{Boisseau_2013}. Fluid flow has also been used to trigger one-off snap-through in applications to on/off switches, valves, or flow regulators. \citet{Gomez2017_2} used a small-scale low-Reynolds-number channel flow to study the one-off snap-through of a buckled sheet embedded at the bottom of a channel, whereas \citet{Arena2017} devised a shape-adapted air inlet with a buckled sheet in which the flow is regulated by the snap-through and snap-back motions at the inlet. \citet{Peretz2020} proposed a slender elastic membrane to make a continuous multi-stable structure in which two different equilibrium states coexist, with a transition region between them.

A buckled sheet undergoes periodic snap-through oscillations under a uniform external flow at high Reynolds numbers. For a single buckled sheet, \citet{HyeonseongJFM2021} identified the conditions for transition between the static equilibrium and snap-through oscillations, and revealed several salient features of the snap-through oscillations such as divergence instability and high bending-energy generation. Furthermore, the snapping motions of tandem buckled sheets under fluid flow have been examined to determine the effects of the upstream buckled sheet on the critical condition and post-critical dynamics of the downstream buckled sheet \citep{JunsooKim2021}. For triboelectric energy harvesting, \citet{HyeonseongNano2020} introduced a buckled elastic sheet of a finite height between two parallel confining walls, where the snap-through oscillations of the sheet by uniform fluid flow in three-dimensional space induce periodic contact and separation with the walls, and showed that this configuration had several regular contact modes depending on the flow speed and gap distance between the walls.

The snap-through phenomenon has only recently been introduced as a method for triboelectric energy harvesting. The aforementioned previous studies approached the problem with experimental measurements, which lack detailed information about flow characteristics around the sheet, contact force, and vortex dynamics in the wake. These aspects are essential to comprehensively understand the fluid--structure interaction principles of triboelectric energy harvesting applications, and can be unraveled using numerical simulations. Therefore, in this study, we numerically investigate the snap-through dynamics of a buckled sheet between two confining walls to elucidate the complicated contact and separation process of the sheet and confining walls. The simulations of snap-through oscillations and periodic contacts with the walls are conducted for the first time to the best of our knowledge. A two-dimensional sheet model and fluid domain are adopted to simplify the problem. The streamwise distance between the two clamped ends of the sheet, the crosswise distance between the two confining walls, the bending stiffness of the sheet, and the free-stream velocity are varied.

Details of model configuration and input parameters, as well as governing equations for numerical simulations, are provided in \S\ref{sec:Problem description}. The coupling of fluid and structure solvers and contact algorithm are described in \S\ref{sec:Numerical method}. The shape of the sheet in equilibrium state and the critical condition for the onset of snap-through are discussed in \S\ref{subSec:Equilibrium state and critical UStar}. In \S\ref{subsec:Post-equilibrium}, the post-critical dynamics of the sheet are analysed, with a particular focus on the contact force and oscillation frequency of the sheet. The effects of the confining walls on flow structures in the post-equilibrium state are then addressed in \S\ref{subSec:Flow structure}. Finally, our findings are summarised in \S\ref{sec:Concluding remarks}.

\section{Problem description}
\label{sec:Problem description}
\subsection{Model and parameters}
\label{subsec:Model and parameters}
\begin{figure}
\centering
\includegraphics{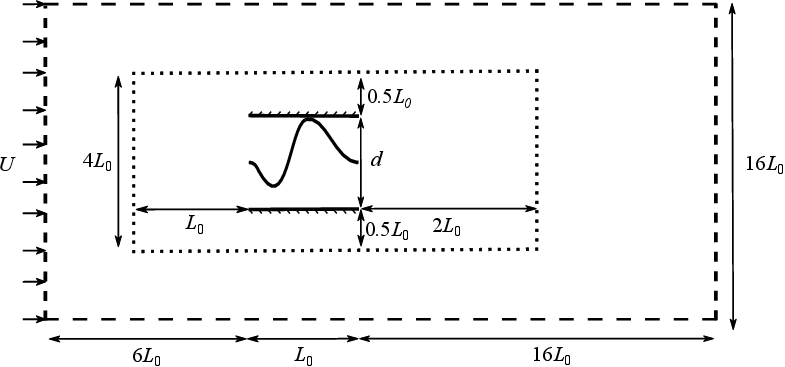}
\caption{Schematic of a buckled sheet between two horizontal confining walls and a fluid domain. The inner dotted rectangle and outer dashed rectangle indicate the region with fine grid resolution and the entire fluid domain, respectively. }
\label{fig_schematic}
\end{figure}

An elastic sheet of length $L$, thickness $h$, density $\rho_s$, and bending stiffness per unit depth $EI(=Eh^3/12)$ is located between two horizontal walls of gap distance $d$ (figure~\ref{fig_schematic}). The distance $L_0$ between the two clamped ends of the sheet is identical to the length of the walls, being smaller than $L$; a bi-stable buckled sheet is formed in the absence of fluid flow. The sheet is exposed to a uniform fluid flow of velocity $U$, density $\rho_f$, and kinematic viscosity $\nu$. The sheet is initially at equilibrium without the fluid flow, and the geometric parameters determine whether the sheet is in contact with the confining wall or not. Although two up--down symmetric shapes of the buckled sheet are possible in equilibrium state without the fluid flow, only one configuration is considered as an initial condition (figure~\ref{fig_schematic}). When the free-stream velocity $U$ surpasses a certain critical value $U_c$, the buckled sheet becomes unstable and snaps to the other side, leading to periodic oscillations \citep{HyeonseongNano2020,HyeonseongJFM2021}.

Instead of finite walls with length $L_0$, infinite walls along the $x$-direction could be considered alternatively: that is, a buckled sheet inside an infinitely long channel of Poiseuille flow. In such a configuration, the incoming flow cannot be diverted at the inlet of the channel, which leads to the significant increase in the pressure force acting on the sheet and the consequent reduction in the critical velocity $U_c$, particularly for cases with large blockage by the sheet inside the channel. The configuration with finite walls is more realistic for actual applications to fluid kinetic energy harvesting from wind and ocean. Hence, the present study adopts the finite walls although the configuration with infinite walls is expected to have some distinct characteristics.

The response of the sheet is affected by several dimensionless parameters: the dimensionless flow velocity ($U^*=U(\rho_f w_0L^2/EI)^{1/2}$), sheet length ratio ($L^*=L_0/L$), gap distance ratio ($d^*=d/L$), blockage ratio ($w_0^*=w_0/d$), mass ratio ($m^*=\rho_s h/\rho_f L$), and Reynolds number ($Re=UL/\nu$). The maximum transverse deflection of the initial buckled sheet along the $y$ direction at $U = 0$ when \emph{no wall} is present is $w_0$, which varies with changes in $L$ or $L_0$. Although $w_0$ is the maximum transverse deflection of the sheet without the walls, it is also used to define the blockage ratio for the cases in contact with the wall. It is because, when the sheet is in contact with the wall, the peak-to-peak transverse distance of the deformed sheet at $U=0$ is similar to $w_0$. The choice of the dimensionless parameters, except for gap distance ratio, follows the work of \citet{HyeonseongJFM2021}, in which detailed reasoning on nondimensionalization is provided. $U^*$ is a particular dimensionless parameter introduced by \citet{HyeonseongJFM2021} for a snap-through model, and is different from the dimensionless flow velocity commonly used for fluttering sheet models \citep{Kim2013,Shoele2016,Hyeonseong2019}.

The dimensionless flow velocity $U^*$ ranges from 5.0--28.0. $L^*$ ranges between 0.6--0.9, and is varied by changing $L_0$ while maintaining $L=1$. $d^*(=0.40$--1.00) is considered to include both contact and non-contact conditions, and the cases without confining walls ($d^*=\infty$) are also considered. $w_0^*$ was originally introduced by \citet{Gomez2017_2} to indicate how much the cross section of the channel is blocked by the buckled sheet. Because $w_0$ is a function of $L$ and $L_0$, $w_0^*$ is not an independent dimensionless parameter, but is determined by $L^*$ and $d^*$. $m^* (=0.75)$ and $Re (=200)$ are fixed, following \citet{HyeonseongJFM2021} who showed that snap-through instability is of divergence type rather than flutter type and these parameters are relatively of less importance in determining critical conditions, compared with $U^*$ and $L^*$. For this reason, although $m^*$ and $Re$ affect post-critical dynamics of the sheet, only $U^*$, $L^*$, and $d^*$ are considered as dimensionless variables in this study. The input parameters of this study are summarised in table~\ref{tab:variables}.

\begin{table*}\caption{Dimensionless input parameters.}\label{tab:variables}
\begin{center}
\begin{tabular}{ c c c }
Parameter & Definition & Range \\[2ex]
$U^*$ & $U(\rho_f w_0L^2/EI)^{1/2}$ & 5.0--28.0
\\
$L^*$ & $L_0/L$ & 0.6--0.9
\\
$d^*$ & $d/L$ & 0.40--1.00
\\
$w_0^*$ & $w_0/d$ & 0.32--0.87
\\
$m^*$ & $\rho_s h/\rho_f L$ & 0.75
\\
$Re$ & $UL/\nu$ & 200
\end{tabular}
\end{center}
\end{table*}

\subsection{Governing equations}
\label{subsec:Governing equations}

For numerical simulations, the non-linear elastic model with an inextensibility constraint, which is commonly used for elastic sheets~\citep{ZhuPeskin2002,CONNELL2007,Tian_2011,Mazharmanesh2020}, is adopted along with the direct-forcing immersed boundary method (IBM) \citep{Uhlmann2005}. The numerical method is implemented in the OpenFOAM framework by developing a new library for the elastic sheet model and IBM. The flow is two-dimensional, incompressible, and laminar, and is governed by the following dimensionless continuity and momentum equations:
\begin{subequations}
\begin{align}
\label{eq_continuity}
\nabla\cdot\boldsymbol{u}&=0, \\
\label{eq_Momentum}
\frac{\partial \boldsymbol{u} }{\partial{t}}+\boldsymbol{u}\cdot\nabla\boldsymbol{u} &=
-\nabla p + \frac{1}{Re} \nabla^2\boldsymbol{u}+\boldsymbol{f},
\end{align}
\end{subequations}
where $\boldsymbol{u}=(u,v)$ and $p$ are the velocity vector and pressure. In \eqref{eq_Momentum}, $\boldsymbol{f}$ is a source term that enforces the continuity of the flow passing the immersed boundary \citep{Uhlmann2005}. The entire fluid domain extends $23 L_0$ and $16 L_0$ in the streamwise and crosswise directions, to minimise the effect of artificial boundary conditions (figure~\ref{fig_schematic}). A uniform flow is applied to the inlet on the left side of the domain. Symmetric conditions are assigned on the top and bottom boundaries, and a zero velocity gradient and constant pressure are imposed at the outlet on the right, with the no-slip condition enforced on the two confining walls.

The two-dimensional equation for the elastic sheet is written in dimensionless form as \citep{CONNELL2007}
\begin{equation}
\label{eq_sheet}
m^*\frac{\partial^2\boldsymbol{X}}{\partial t^2}
-\frac{\partial}{\partial s}\left(T(s)\frac{\partial\boldsymbol{X}}{\partial s}\right)
+k_b\frac{\partial^4\boldsymbol{X}}{\partial s^4}=\boldsymbol{F}_f+\boldsymbol{F}_c+m^* Fr\frac{\boldsymbol{g}}{g},
\end{equation}
where $m^*(=\rho_s h/\rho_f L)$ is the mass ratio, $\boldsymbol{X}=(x,y)$ is the position vector, $s$ is the curvilinear coordinate starting from the left end of the sheet, and $k_b=EI/\left(\rho_fU^2 L^3\right)$ is the bending coefficient. $\boldsymbol{F}_f$ is the fluid force exerted on the sheet, which is calculated by extrapolating the pressure and stress on the structure surface, $\boldsymbol{F}_c$ is the force exerted by the wall contact, and $Fr=gL/U^2$ is the Froude number. $T$ is the tension force, defined as
\begin{equation}
\label{eq_tension}
T(s)=k_s
\left[
\left(
\frac{\partial\boldsymbol{X}}{\partial s} \cdot \frac{\partial\boldsymbol{X}}{\partial s}
\right)^{1/2}-1
\right],
\end{equation}
where $k_s=200$ is a sufficiently large coefficient chosen to ensure the inextensibility of the sheet; we confirmed that our simulations were insensitive to the value of this parameter. The last term in \eqref{eq_sheet} for gravitational effect is only considered for validation cases in \S\ref{subsec:Convergence test and validation}. At both clamped ends of the sheet, the boundary conditions are
\begin{subequations}
\begin{align}
\boldsymbol{X}=(0,0)\quad \textrm{and} \quad \frac{\partial{\boldsymbol{X}}}{\partial{s}}=(1,0)\quad \textrm{at} \quad s=0,\label{BC1}\\
\boldsymbol{X}=(L_0,0)\quad \textrm{and} \quad \frac{\partial{\boldsymbol{X}}}{\partial{s}}=(1,0) \quad \textrm{at} \quad s=1.\label{BC2}
\end{align}
\label{BC1_2}
\end{subequations}

The equilibrium shape of the sheet in the absence of flow is acquired by numerically solving \eqref{eq_sheet}. Depending on the sheet length ratio $L^*$ and the gap distance ratio $d^*$, the sheet can be in contact with the wall or separated from the wall. For non-contact cases, the exact solution for the equilibrium shape at zero flow velocity is obtained from the solution in \citet{Beharic2014}. To find the equilibrium shape at zero flow velocity for all contact cases, first we place the walls at $d^{*}=0.7$ where contact does not occur, so that we can use the exact solution of a sheet with no contact. The walls are then gradually moved closer together until a specified $d^*$ reaches.

\section{Numerical method}
\label{sec:Numerical method}
\subsection{Fluid and structure solvers}
\label{subsec:Fluid and structure solvers}
Equations~\eqref{eq_continuity} and \eqref{eq_Momentum} are solved using the PISO algorithm as described by \citet{Jasak}, with some modifications to consider the forcing term $\boldsymbol{f}$. All terms are discretised by second-order Gauss finite-volume integration, except for the convection term discretised by the second-order upwind scheme. The implicit first-order Euler method is used for time marching. To find $\boldsymbol{f}$, a momentum equation without $\boldsymbol{f}$ is first solved to obtain $\tilde{\boldsymbol{u}}$:
\begin{equation}
\label{eq_Momentum_Utilde}
\frac{\partial \tilde{\boldsymbol{u}} }{\partial{t}}+\tilde{\boldsymbol{u}}\cdot\nabla\tilde{\boldsymbol{u}}=
-\nabla p + \frac{1}{Re} \nabla^2\tilde{\boldsymbol{u}}.
\end{equation}
$\tilde{u}$ is an intermediate velocity, which is different from the solution of the previous time step, and it is not the exact solution of the current time step. $\tilde{u}$ is then interpolated to the Lagrangian grid for the sheet:
\begin{equation}
\label{eq_u_to_U}
\tilde{\boldsymbol{U}}=\int_{\Omega_f} \tilde{\boldsymbol{u}}\delta(\boldsymbol{X}-\boldsymbol{x}) \text{d}\boldsymbol{x},
\end{equation}
where $\delta$ is a function defined by \citet{Roma1999} as 
\begin{equation}
\label{DeltaFunction}
\delta(\boldsymbol{X}-\boldsymbol{x})=\frac{1}{\Delta x_r^2}\phi\left(\frac{X-x}{\Delta x_r}\right)\phi\left(\frac{Y-y}{\Delta x_r}\right),
\end{equation}
where $\Delta x_r$ is the size of uniform grids in the fluid region around the sheet in both $x$ and $y$ directions. $\phi(r)$ is a continuous function defined as
\begin{equation}
\label{phiInDelta}
    \phi(r)=\left\{
                \begin{array}{ll}
                  \frac{1}{6}\left(5-3|r|-\sqrt{-3(1-|r|)^2+1}\right)\quad &0.5\leq|r|\leq 1.5,\\
                  \frac{1}{3}\left(1+\sqrt{-3r^2+1}\right)\quad &|r|\leq0.5,\\
                  0\quad &\text{otherwise.}
                \end{array}
              \right.\end{equation}
The difference between the intermediate velocity $\tilde{\boldsymbol{U}}$ of the sheet, which is interpolated from the fluid, and the velocity $\boldsymbol{U}$ of the sheet in the previous time step, which is provided by the structure solver, is used to define a forcing term on the Lagrangian points~\citep{Uhlmann2005}:
\begin{equation}
\label{eq_langrangian_forcing}
\boldsymbol{F}=\frac{\text{d}(\boldsymbol{U}-\tilde{\boldsymbol{U}})}{\text{d}t}.
\end{equation}
$\boldsymbol{f}$ is then determined by transferring $\boldsymbol{F}$ to the Eulerian domain for the flow field:
\begin{equation}
\label{eq_F_to_f}
\boldsymbol{f}=\int_{\Omega_s} \boldsymbol{F}\delta(\boldsymbol{x}-\boldsymbol{X}) \text{d}s.
\end{equation}
The divergence-free velocity is then given by solving for the pressure.

The inextensibility condition of the elastic sheet is a well-known issue in the coupling of a fluid and an elastic sheet \citep{Huang2007}. To avoid instability in the numerical simulations and large errors in the length change, the time step must be much smaller than the value specified by the Courant--Friedrichs--Lewy condition \citep{Huang2007,Shoele2016,Ryu2018,Mazharmanesh2020}. Here, sub-cycling is suggested to solve \eqref{eq_sheet}, as this considerably reduces the computation time while accurately preserving the sheet length. Equation~\eqref{eq_sheet} for the sheet is solved implicitly at the end of the PISO algorithm with the fluid pressure and velocity at time $t = (n+1)\Delta t$. For $N$ sub-cycles of the structure solver with the sub-cycle time step $\Delta\tau=\Delta t/N$,
\begin{equation}
\label{eq_sheet_m}
m^*\frac{\boldsymbol{X}^{m+1}-2\boldsymbol{X}^m+\boldsymbol{X}^{m-1}}{\Delta \tau}
-\frac{\partial}{\partial s}\left(T^m(s)\frac{\partial\boldsymbol{X}^{m}}{\partial s}\right)
+k_b\frac{\partial^4\boldsymbol{X}^{m}}{\partial s^4}=\boldsymbol{F}_f^{n+1}+\boldsymbol{F}_c^m+m^* Fr\frac{\boldsymbol{g}}{g},
\end{equation}
where the superscript $m$ indicates the $m$-th sub-cycle time step. After sub-cycling, the location of the sheet is obtained at time $t = (n+1)\Delta t$. The spatial discretisation of \eqref{eq_sheet_m} is performed using the central differencing method and on staggered grids of the same size as the uniform fluid grids around the sheet.

The collision of the sheet with a confining wall is implemented using an artificial repulsive force $\boldsymbol{F}_c$, which prevents the sheet from penetrating the wall. The total repulsive force applied on each element $i$ of the sheet is calculated as a summation over the forces exerted by all elements of the wall \citep{Glowinski2001}.
\begin{equation}
\label{contact_force}
\boldsymbol{F}_{c,i}=\begin{cases}
\sum\limits_{j=1}^{k} \frac{1}{\epsilon}\boldsymbol{d}_{ij}\delta(\boldsymbol{d}_{ij})\left(\xi-|\boldsymbol{d}_{ij}|\right)^2 & \text{if } |\boldsymbol{d}_{ij}| \leq \xi,\\
0 & \text{otherwise},
\end{cases}
\end{equation}
where $\boldsymbol{d}_{ij}=\boldsymbol{X}_i-\boldsymbol{X}_j$ is the distance vector between the centres of element $i$ of the sheet and element $j$ of the wall. $\xi=2 \Delta x_r$ is the reference distance within which the force is active; $\Delta x_r$ is the grid size of the fluid around the sheet. $\epsilon$ is a parameter that regulates the force to prevent penetration through boundaries. In all simulations, $\epsilon=\Delta x_r$ was found to be small enough to avoid penetration.

The numerical procedure for coupling the fluid and structure solvers is as follows.
\begin{enumerate}
\item Solve \eqref{eq_Momentum_Utilde} for the intermediate flow velocity $\tilde{\boldsymbol{u}}$;
\item Find a source term $\boldsymbol{f}$ from \eqref{eq_u_to_U}--\eqref{eq_F_to_f};
\item Solve with the PISO algorithm to obtain $p^{n+1}$ and $\boldsymbol{u}^{n+1}$;
\item Calculate the fluid force on the sheet $\boldsymbol{F}_f^{n+1}$;
\item Solve \eqref{eq_sheet_m} for $N$ sub-cycles to obtain the sheet location $\boldsymbol{X}^{n+1}$ and velocity $\boldsymbol{U}^{n+1}$.
\end{enumerate}

Snap-through is accompanied by the rapid and complex shape morphing that may cause instability in both the flow field and the structure, and sometimes the failure of the simulation. In our model, the instability occurs as the saw shape of the sheet and the checkered pressure field around the sheet. To overcome these numerical issues, a small time step of $\Delta t=0.0005$ is chosen for the fluid solver in all cases, and a time step for the structure solver is set to be $25$ times smaller ($N=25$). We confirmed that these time steps were sufficiently small to ensure convergence. Larger time steps for the fluid solver cause checkered pressure fields and the saw shape of the sheet, while fewer iterations for the structure solver lead to its failure.

\subsection{Grid convergence test and validation}
\label{subsec:Convergence test and validation}

For the fluid domain, uniform grids of size $\Delta x_r=L/150$ are used around the sheet and confining walls inside the dotted rectangular region in figure~\ref{fig_schematic}. Outside this region, the grids gradually coarsen so that the largest spacing at the domain boundaries is $\Delta x=\Delta x_r/4$. A value of $\Delta x_r=L/150$ was chosen after conducting extensive grid convergence tests for four grid layouts with $\Delta x_r=L/60$, $L/100$, $L/150$, and $L/225$, with particular focus on the contact force averaged over the time span of $t=20.0$--60.0 and the oscillation frequency as they are important parameters in this study. Cases with no contact or weak contact are hardly affected by the gird size. The general response of the sheet and the flow field are the same for $L/100$ and finer grid layouts. However, in few cases, we observed large variations in contact force, which also causes considerable errors in oscillation frequency. Such variations are negligible between the selected grid layout of $\Delta x_r=L/150$ and the smaller grid layout of $\Delta x_r=L/225$. For example, in the case of $L^*=0.6$, $d^*=0.40$, and $U^*=22.0$, the variations in the averaged contact force are 4.3\%, 2.4\%, and 0.5\% for $\Delta x_r=L/60$, $L/100$, and $L/150$, respectively, compared with the finest grid layout of $\Delta x_r=L/225$.

\begin{figure}
\centering
\includegraphics{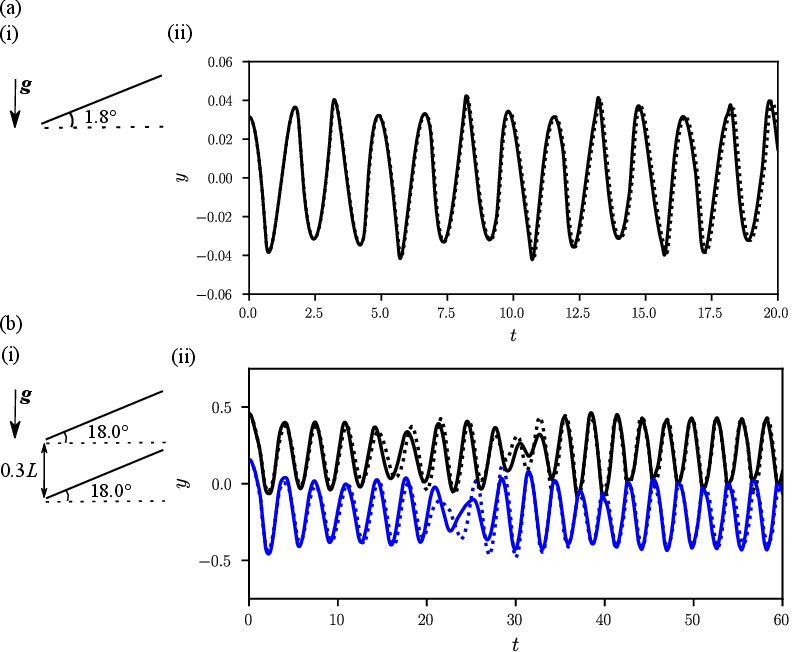}
\caption{(\emph{a}i) Schematic of a single inclined sheet exposed to the gravitational force and (\emph{a}ii) comparison of tip displacement of the sheet between our numerical solution (solid line) and analytical solution (dotted line).
(\emph{b}i) Schematic of two side-by-side sheets exposed to an inflow of velocity $U$ and gravity and (\emph{b}ii) comparison of tip displacements of the upper (black) and lower (blue) sheets between our numerical solution (solid lines) and \citet{Huang2007} (dotted lines). }
\label{validation}
\end{figure}

The validation of the structure solver is first performed using an elastic sheet in the absence of fluid. For the validation, the right end of the sheet is assigned to be free with the following boundary conditions:
\begin{equation}
T=0, \quad \frac{\partial^2{\boldsymbol{X}}}{\partial^2{s}}=(0,0),\quad \frac{\partial^3{\boldsymbol{X}}}{\partial^3{s}}=(0,0)\quad \textrm{at} \quad s=1.
\end{equation}
The sheet is initially straight with an initial inclination angle of $\theta=1.8^\circ$ with respect to the $x$-axis, and is then exposed to the gravitational force (figure~\ref{validation}\emph{a}i). For $L=1$, $Fr=10$, $k_b=0$, $k_s=1000$, and $50$ Lagrangian cells along the sheet, the time history of the tip displacement in our simulation is in excellent agreement with that of the analytical solution given by \citet{Huang2007} (figure~\ref{validation}\emph{a}ii).

To validate the coupling of the fluid and structure solvers and the contact algorithm, two side-by-side sheets with a pole-to-pole distance of $D=0.3L$ along the $y$-axis are considered (figure~\ref{validation}\emph{b}i). Both sheets have the same properties of $L =1$, $m^*=1.5$, $Fr=0.5$, $k_b=0.0015$, and $k_s=100$ with an initial inclination angle of $\theta=18.0^\circ$. They are immersed inside a fluid domain extending over $-2\leq x \leq 6$ and $-4\leq y \leq 4$ and are subjected to a uniform fluid flow at $Re=200$. A grid spacing of $L/64$ is used for both the structure and fluid domains. The two sheets start in-phase oscillations, and at a certain time they commence out-of-phase oscillations, causing multiple collisions between them. The contact force between the two sheets is modeled in the same way as that between the sheet and the wall in our snap-through model, using~\eqref{contact_force}. The displacements of the free right ends of both sheets are in good agreement with those of \citet{Huang2007} (figure~\ref{validation}\emph{b}ii). Slight discrepancies are observed in $t=15$--35 when a transition from in-phase to out-of-phase oscillations occurs, because the process of transition to out-of-phase oscillations may differ by the numerical method employed. In out-of-phase oscillations after $t = 35$, the results of the two methods become similar again.

\subsection{Dynamic mode decomposition}
\label{subsec:Dynamic mode decomposition}
Dynamic mode decomposition (DMD) is a method for extracting the dominant and coherent modes of a dynamical system \citep{Schmid2010}. The non-linear behaviour of the dynamical system can be examined through a linear approximation if a sufficiently short time interval is considered.
\begin{equation}
\label{DMDLinear}
\frac{\mathrm{d}\boldsymbol{\Phi}}{\mathrm{d}t}=\mathcal{A}\boldsymbol{\Phi},
\end{equation}
where $\boldsymbol{\Phi}(t)\in\mathbb{R}^n$ is an $n$-dimensional vector representing the state of the system at time $t$. From a total of $m$ measurements of states $\boldsymbol{\Phi}_k$ at equally spaced time intervals $\delta t$, with the initial state of $\boldsymbol{\Phi}(0)$, the solution of \eqref{DMDLinear} can be expressed as \citep{Kutz2016}
\begin{equation}
\label{DMDSolution}
\boldsymbol{\Phi}(t)=\sum\limits_{k=1}^{n} \boldsymbol{\psi}_k \exp(\omega_k t)b_k
\end{equation}
The solution of $\boldsymbol{\Phi}(t)$ requires the eigenvectors $\boldsymbol{\psi}_k$ and eigenvalues $\omega_k$ of the matrix $\mathcal{A}$ and the coefficients $b_k$. These terms are extracted by considering a discretised representation of \eqref{DMDLinear} as
\begin{equation}
\label{DMDDiscreet}
\boldsymbol{\Phi}_{k+1}=\boldsymbol{A\Phi}_{k},
\end{equation}
with $\boldsymbol{A}=\exp(\mathcal{A}\delta t)$. We follow the DMD algorithm explained in \citet{Kutz2016} to solve \eqref{DMDDiscreet} using a low-rank eigendecomposition of $\boldsymbol{A}$ and find the parameters in \eqref{DMDSolution}.

The real and imaginary parts of $\omega_k=2\pi f_k+\mathrm{i}\zeta_k$ indicate the oscillation frequency $f_k$ and growth/decay rate $\zeta_k$ of each mode, respectively. For the snap-through oscillations of the sheet, we make this parameter dimensionless using twice the maximum transverse deflection of the unbounded buckled sheet without flow ($2w_0$) as the characteristic amplitude \citep{HyeonseongJFM2021}: $\omega_k^*=\omega_k(2w_0)/U$. The snap-through frequency is then defined to be equal to the frequency of the first dominant oscillatory DMD mode ($f^*=f_d(2w_0)/U$). DMD is applied to snapshots of the sheet over the time span $t=20.0$--60.0 at intervals of $\delta t=0.008$, which is sufficient to accurately capture the dominant frequency. The same time span is used to extract the DMD modes of the vorticity field. The DMD modes of the vorticity field are obtained from snapshots of the rectangular fluid domain $[-0.2,8.0]\times [-1.0,1.0]$.

\section{Results and discussion}
\label{sec:Results and discussion}

\subsection{Shape in equilibrium state and critical flow velocity}
\label{subSec:Equilibrium state and critical UStar}

\begin{figure}
\centering
\includegraphics{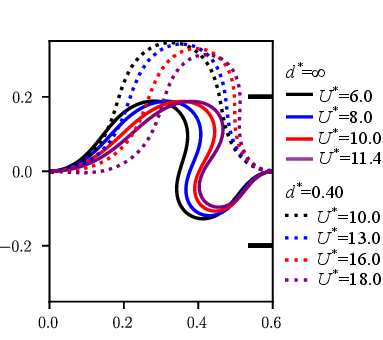}
\caption{Equilibrium shapes of the sheet for several dimensionless flow velocities $U^*$. The horizontal black solid bars on the right end of the panel denote the positions of the confining walls for $d^* = 0.40$. The critical flow velocity $U^*_c$ is 11.6 for $d^* = \infty$ and 18.3 for $d^*=0.40$.}
\label{EqShapeVariationWithU}
\end{figure}

When the dimensionless flow velocity $U^*$ increases from zero, the buckled sheet maintains a quasi-static equilibrium shape up to a certain critical condition. Even at the pre-critical condition, the sheet may snap a few times due to the sudden rise of the fluid force at the beginning of the simulation. However, the snapping motion does not persist and the sheet reaches a stable equilibrium on either the upper or lower side of the channel centreline (the horizontal line that connects the two ends of the sheet). In this section, if the equilibrium state of the sheet is on the lower side, it is mirrored to the upper side for ease of comparison. The equilibrium shapes of the sheet differ between contact and non-contact cases (figure~\ref{EqShapeVariationWithU}). The smallest sheet length ratio $L^*(=0.6)$, which initially has the largest transverse deflection in the absence of the confining walls, and the smallest gap distance ratio $d^*(=0.40)$ are shown in figure~\ref{EqShapeVariationWithU} to illustrate the dramatic effects of sheet--wall interactions. As $U^*$ increases, the stable sheet gradually leans along the streamwise direction until the sheet can no longer maintain the equilibrium and snaps to the other side of the channel at the critical velocity $U^* = U^*_c$.

The unbounded case without the confining walls has one apex above the centreline near the midpoint ($x=0.3$), which gradually shifts backwards (along the $x$-axis) with increasing $U^*$. When $U^*$ is close to $U^*_c$, the front (left) part of the sheet crosses the centreline slightly, having a negative $y$ value in figure~\ref{EqShapeVariationWithU}. By contrast, for $d^* =0.40$ in figure~\ref{EqShapeVariationWithU}, the sheet is highly deformed, making contact with the upper wall. It has one nadir below the centreline on the rear (right) part of the sheet and one apex above the centreline on the left of the nadir. Thus, the sheet blocks a significant portion of the channel, which hinders the fluid from passing through the channel and delays the onset of periodic snap-through. Compared to a channel of the same gap distance ($d^* =0.40$) without the sheet, the flow rate inside the channel is reduced about 10 times from 0.310 to 0.036 at $U^* = 10.0$ when a sheet of $L^* =0.6$ is located inside the channel. By gradual morphing of the sheet, the flow rate through the channel becomes greater with increasing $U^*$, and it amounts to be 0.065 at $U^*=18.0$ before the occurrence of instability, which is still much less than the flow rate without the sheet.

\begin{figure}
\centering
\includegraphics{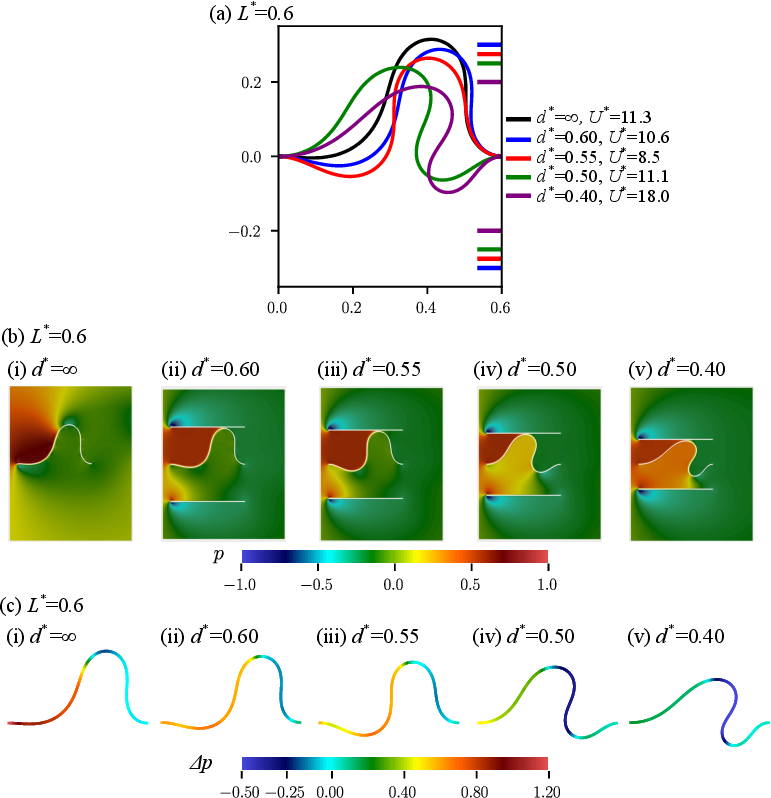}
\caption{(\emph{a}) Equilibrium shapes of the sheet just before the transition to periodic snap-through for five confining wall distances ($L^*=0.6$). The horizontal colour bars on the right end of the panel denote the positions of the confining walls for each colour. (\emph{b}) Contours of normalised fluid pressure $p$. (\emph{c}) Distribution of net pressure $\Delta_p(=p_{u}-p_{l})$ acting on the sheet. The colour on the sheet indicates the value of net pressure.
}\label{sheet_equilibrium_LStar_06}
\end{figure}

Figure~\ref{sheet_equilibrium_LStar_06}(\emph{a}) presents the equilibrium shapes of the sheet with $L^*=0.6$ when $U^*$ is slightly less than the critical value for five gap distances, $d^*=0.40$--0.60 and $d^* = \infty$ (no confining walls); note that the $U^*$ values differ in the five cases. The corresponding pressure fields around the sheet, which are normalised by $\rho_f U^2$ (figure~\ref{sheet_equilibrium_LStar_06}\emph{b}), and the net pressure $\Delta p$ acting on the sheet (figure~\ref{sheet_equilibrium_LStar_06}\emph{c}) are also depicted for the five gap distances. $\Delta p(=p_{u}-p_{l})$ is the difference in the normalised pressure between the upper and lower surfaces of the sheet.

For the case of $d^*=0.60$, the apex height is slightly lower because the sheet is constricted by the wall, and the front part of the sheet crosses farther below the centreline compared with the unbounded case of $d^* = \infty$. Although the net pressure force on the front part of the sheet for $d^*= 0.60$ is lower than for $d^* = \infty$ (figure~\ref{sheet_equilibrium_LStar_06}\emph{c}i,\emph{c}ii), the instability can occur at a lower dimensionless velocity: $U^*_c$ = 11.5 for $d^* = \infty$ and 10.8 for $d^* = 0.60$. When the confining walls are placed more closely together at $d^*=0.55$, $U^*_c$ drops dramatically to 8.7. Despite minor change in the pressure field of the flow and the pressure force on the sheet (figure~\ref{sheet_equilibrium_LStar_06}\emph{b}iii,\emph{c}iii), there are distinct differences in the shape of the sheet between $d^*=0.60$ and 0.55. Therefore, the reduction in the critical velocity is attributed to the shape change of the sheet resulting from stronger contact with the confining wall. While the front part of the sheet slightly crosses the centreline for $d^* = \infty$, it notably passes the centreline and forms a nadir for $d^*=0.55$ (figure~\ref{sheet_equilibrium_LStar_06}\emph{a}). Because of this particular configuration, the sheet is most susceptible to instability when $d^*=0.55$, and the transition to periodic snap-through occurs at lower $U^*$ than for the other cases.

For contact cases, the initial shape of the sheet has a nadir on its left and an apex on its right at $U^* = 0$. For $L^* = 0.6$, the smallest gap distance ratio at which the sheet is able to preserve this configuration under the fluid flow before the transition to the post-equilibrium state is $d^*=0.55$. Further reducing the gap distance ratio from $d^*=0.55$ causes the sheet to become temporarily unstable at a certain flow velocity and snap to the other side due to the forces imposed by the flow and wall contact, accompanied by a dramatic shift in the equilibrium shape. The new equilibrium shape, which is mirrored in figure~\ref{sheet_equilibrium_LStar_06}(\emph{a}) for ease of comparison, is slanted in the streamwise direction, yielding a deep nadir and the highly curved deformation of an S-shape on the right of the sheet. Such a shift in the equilibrium shape was also observed in the experimental study of \citet{HyeonseongNano2020} for the length ratio of $L^*=0.75$. They reported that the apex occurred at the front of the sheet for $d^*<0.42$ and at the rear for $d^*>0.42$. For $d^*=0.42$, the sheet had two equilibrium shapes with different $U^*_c$ values, one with the nadir on the rear (higher $U^*_c$) and the other on the front (lower $U^*_c$). The multiple (two) equilibrium shapes at a specific $d^*$ is not observed in our numerical simulations for several $d^*$ values considered in the present study. This indicates that $d^*=0.55$ is not the exact threshold for the shift in shape and two equilibrium shapes may exist at a specific $d^*$ which is not covered in the current simulations. Except for $L^*=0.9$ which produces no contact with the confining wall, a shift in the equilibrium shape is also observed for $L^*=0.7$ and $0.8$ at a certain flow velocity by reducing $d^*$ from 0.50 to 0.45 and from 0.45 to 0.40, respectively (table~\ref{tab:shapeshift_dStar}). Interestingly, the shape shift arises consistently when the blockage ratio $w_0^*$ exceeds a threshold value in the narrow range of 0.66--0.70. This suggests that the blockage ratio is the dominant geometric parameter in determining the shift in the equilibrium shape.

\begin{table*}\caption{Gap distance ratio $d^*$ and blockage ratio $w_0^*$ at which the shift in equilibrium shape occurs at a certain flow velocity, moving the nadir from the left part to the right part of the sheet.}\label{tab:shapeshift_dStar}
\begin{center}
\begin{tabular}{ c c c }
$L^*$ & Nadir on the left & Nadir on the right \\ [2ex]
0.6 & $d^*$=0.55, $w_0^*$=0.63 & $d^*$=0.50, $w_0^*$=0.69 \\
0.7 & $d^*$=0.50, $w_0^*$=0.63 & $d^*$=0.45, $w_0^*$=0.70 \\
0.8 & $d^*$=0.45, $w_0^*$=0.59 & $d^*$=0.40, $w_0^*$=0.66 \\
\end{tabular}
\end{center}
\end{table*}

The shifted equilibrium shapes of $d^*=0.50$ and $0.40$ ($L^*=0.6$) alter the surrounding flow field remarkably and cause an increase in the pressure of the fluid entrained below the apex (figure~\ref{sheet_equilibrium_LStar_06}\emph{b}iv,\emph{b}v). Because a large portion of the channel is blocked ($w_0^*=0.69$ and 0.87 for $d^*=0.50$ and $0.40$, respectively), the fluid flow is hindered from passing through the gap between the sheet and the confining wall, which raises the pressure of the entrained flow below the apex. The inclination of the sheet along the streamwise direction due to confinement by the confining walls also contribute to greater pressure in the entrained flow. The pressure increase below the apex weakens the net pressure $\Delta p (= p_{u}-p_{l})$ imposed on the front part of the sheet, but strengthens the net pressure on its rear part (figure~\ref{sheet_equilibrium_LStar_06}\emph{c}iv,\emph{c}v). Along with the formation of the nadir on the rear part of the sheet, this change in the distribution of the net pressure force makes it difficult for the sheet to snap, leading to an increase in the critical velocity from $U^*_c = 8.7$ for $d^*=0.55$ to 11.3 for $d^*=0.50$. The change in pressure distribution is more pronounced for $d^*=0.40$. Compared with $d^*=0.50$, the greater magnitude of net pressure $\Delta p$ and the formation of the deeper nadir on the rear of the sheet result in a significant increase in $U^*_c$ to 18.3. That is, after the occurrence of the shape shift, the net pressure force on the sheet, which itself is strongly affected by the blockage of the channel, becomes an important factor in determining the critical flow velocity.

\begin{figure}
\centering
\includegraphics{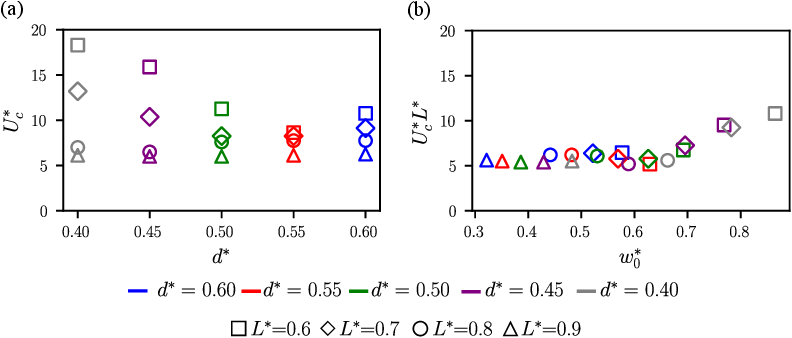}
\caption{(\emph{a}) Dimensionless critical flow velocity $U^*_c$ for the onset of periodic snap-through with respect to gap distance ratio $d^*$ for different length ratios $L^*$. (\emph{b}) $U^*_cL^*$ with respect to blockage ratio $w_0^*$. In (\emph{a}, \emph{b}), the colours of the markers denote $d^*$, and their shapes denote $L^*$.}
\label{UCr}
\end{figure}

As mentioned above, when the sheet is in contact with the confining wall for a given sheet length ratio $L^*$ (except for $L^*=0.9$), the dimensionless critical flow velocity $U^*_c$ decreases with the gap distance ratio $d^*$ up to a certain value, and then increases beyond this threshold. This trend is observed more distinctly for $L^*=0.6$ than for $L^* = 0.7$ and 0.8 (figure~\ref{UCr}\emph{a}). Moreover, according to figure~\ref{UCr}(\emph{a}), $U^*_c$ tends to decrease with $L^*$ for a given gap distance ratio $d^*$, including $d^* = \infty$. The smallest gap distance ratio $d^*=0.40$ produces the largest drop in $U^*_c$, from $18.0$ for $L^*=0.6$ to $6.2$ for $L^*=0.9$. A similar drop in $U^*_c$ is also observed for $d^*=0.45$, but for the other cases, the drop in $U^*$ is much smaller.

In figure~\ref{UCr}(\emph{b}), $U^*_c L^*$ is plotted with respect to the blockage ratio $w_0^*$. Evidently, for $w_0^*<0.66$, $U^*_c L^*$ is between 5.2 and 7.0, implying that $U_c^*$ scales inversely with $L^*$ and is relatively unaffected by $w_0^*$. However, beyond $w_0^*=0.66$, the effect of channel blockage on the critical velocity becomes significant. Earlier, we reported that all cases with $w_0^*\geq 0.66$ undergo a shift in the equilibrium shape and form a deep nadir on the right of the sheet (table~\ref{tab:shapeshift_dStar}). After the shape shift that occurs beyond $w_0^*=0.66$, $L^*U_c^*$ increases almost linearly with $w_0^*$ and reaches a peak value of 10.7 for $w_0^*=0.87$ with the strongest blockage effect.

\subsection{Dynamics in post-equilibrium state}
\label{subsec:Post-equilibrium}

When the free-stream velocity $U^*$ exceeds the critical value of $U^*_c$, the sheet no longer persists quasi-static deformation, but exhibits repeated snap-through oscillations, periodically contacting the confining walls. Snap-through in each cycle could be characterised by rapid release of stored bending energy. This change is accompanied by complex shape morphing that could occur in a time interval as small as 10\% of a cycle or in an interval as large as half of a cycle, depending on the dimensionless velocity, gap distance ratio, and length ratio. The characteristics of bending energy for a snapping sheet were discussed by~\citet{HyeonseongJFM2021}, and will not be examined further in the current study. Here, we primarily focus on the contact force between the sheet and the wall and the oscillation frequency, which are the parameters of interest for triboelectric energy harvesting applications.

\subsubsection{Symmetric and asymmetric oscillations}
\label{subsubSec:asymmetric snapthrough}

\begin{figure}
\centering
\includegraphics{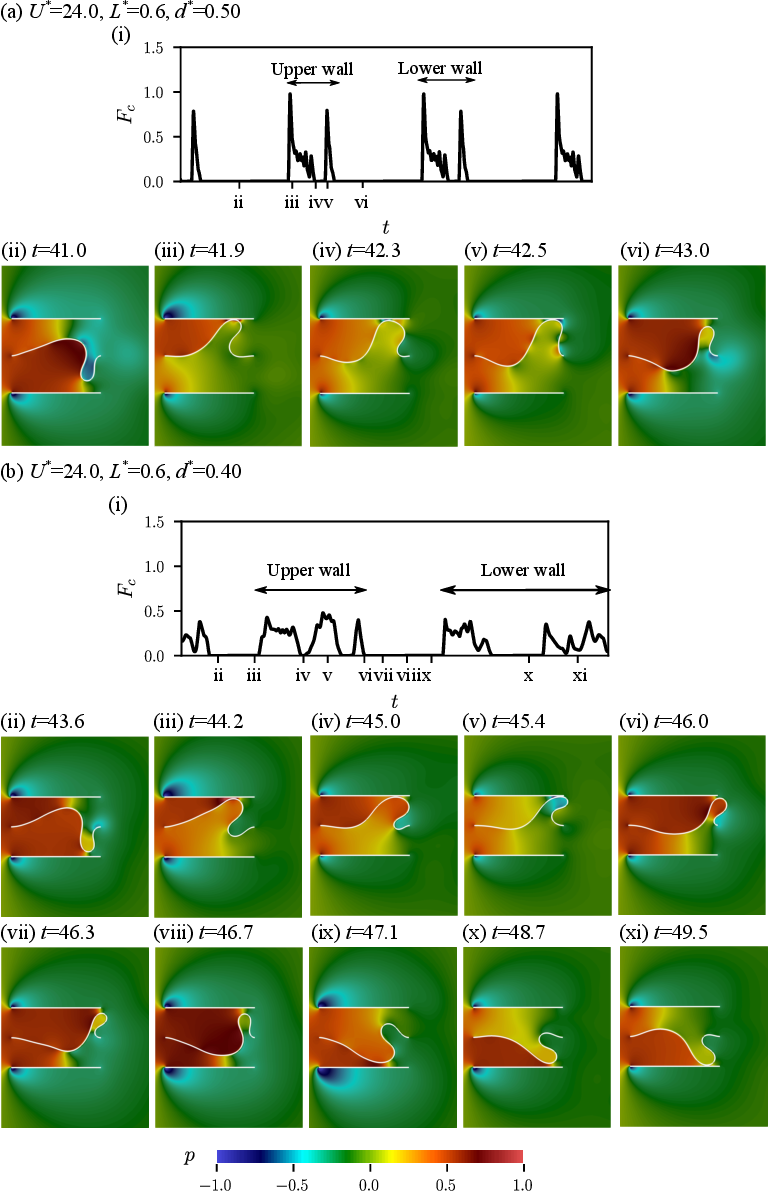}
\caption{(\emph{a}i, \emph{b}i) Temporal variations in contact force $F_c$. In (\emph{a}i) and (\emph{b}i), Roman numbers on the horizontal axis correspond to the sequential snapshots of pressure contours in (\emph{a}ii--\emph{a}vi) and (\emph{b}ii--\emph{b}xi), respectively. For the symmetric case (\emph{a}), only half of the cycle for shape morphing on the upper side of the centreline is illustrated. See supplementary movie 1.}
\label{pressure_asymmetryShapes_d04}
\end{figure}

Before discussing the contact mode and force, we report a particular behaviour of the oscillating sheet that can be attributed to its strict confinement within the channel. Generally, the shape of the oscillating sheet is almost symmetric between the upper and lower sides of the channel centreline. However, by strengthening the sheet--wall interaction (reducing $L^*$ and $d^*$), it is possible to break the symmetry of the oscillations. Snapshots of the pressure field and the magnitudes of the contact force integrated over the sheet ($F_c=|\sum_{i}\boldsymbol{F}_{c,i}|$) are compared between symmetric and asymmetric cases in figure~\ref{pressure_asymmetryShapes_d04} and supplementary movie 1; see~\eqref{contact_force} for the definition of $\boldsymbol{F}_{c,i}$. Here, $F_c$ is the dimensionless contact force normalised by $\rho_f U^2 L$. The case of $L^* = 0.6$, $d^* = 0.50$, and $U^*=24.0$ exemplifies the symmetric behaviour of the sheet, which is evident in the time history of the contact force (figure~\ref{pressure_asymmetryShapes_d04}\emph{a}i). While the sheet approaches a confining wall and first impacts the confining wall with the generation of a peak in $F_c$, the fluid pressure in front of the contact region increases (figure~\ref{pressure_asymmetryShapes_d04}\emph{a}ii,\emph{a}iii). Sequentially, the apex separates from the wall slightly while moving along the streamwise direction, and hits the wall again, while the front part of the sheet moves closer to the centreline (figure~\ref{pressure_asymmetryShapes_d04}\emph{a}iv--\emph{a}vi), which is followed by the next snap-through to the opposite side.

By decreasing $d^*$, the apex of the sheet becomes more displaced in the streamwise direction before snapping to the other side, eventually resulting in the contact force and shape changing substantially from those of the symmetric case. The asymmetric deformation of the sheet near the upper and lower walls is specific to cases with the extreme confinement of $L^*=0.6$ and $d^*=0.40$ and a high velocity range of $U^*=19.0$--28.0, which is exemplified in figure~\ref{pressure_asymmetryShapes_d04}(\emph{b}) for $U^*=24.0$. After contact with the upper wall, the apex of the sheet moves along the streamwise direction, and may even pass the clamped right end at the centreline, yielding a very high curvature on the rear part of the sheet (figure~\ref{pressure_asymmetryShapes_d04}\emph{b}v,\emph{b}vi). This excessive streamwise displacement of the sheet enables an increase in flow velocity on the left of the apex and above the sheet, which leads to a reduction in the fluid pressure above the sheet (figure~\ref{pressure_asymmetryShapes_d04}\emph{b}v). The subsequent deceleration and stagnation of sheet movement in the streamwise direction induces a notable pressure increase above the sheet (figure~\ref{pressure_asymmetryShapes_d04}\emph{b}vi,\emph{b}vii).

The sheet then moves toward the lower wall with a shape clearly different from that of the instant before impacting the upper wall (compare figures~\ref{pressure_asymmetryShapes_d04}\emph{b}ii and~\emph{b}vii). According to figure~\ref{pressure_asymmetryShapes_d04}(\emph{b}i), the sheet slides along the upper wall during the contact process with short separation. By contrast, the sheet undergoes a long separation from the lower wall and a subsequent bouncing behaviour. Several bounces near the lower wall in the transverse direction while moving to the right (figure~\ref{pressure_asymmetryShapes_d04}\emph{b}ix--\emph{b}xi) prevent the fast streamwise sliding that occurs on the upper side of the channel. Consequently, the temporal characteristics of the contact force differ starkly between the contact phases of the upper wall and the lower wall in figure~\ref{pressure_asymmetryShapes_d04}(\emph{b}).

\subsubsection{Contact force}
\label{subsubSec:Contact force}
The possibility of periodic contact with the confining walls and the intensity of the sheet--wall interaction depend on the sheet length ratio $L^*$, gap distance ratio $d^*$, and dimensionless flow velocity $U^*$. For given $L^*$ and $d^*$, a necessary condition for the occurrence of contact in the absence of flow is that the maximum transverse deflection $w_0$ of a sheet without confining walls is greater than $d/2$, corresponding to $w_0^*>0.50$. However, in the presence of flow, the behaviour of the oscillating sheet is influenced by $U^*$, causing substantial changes in the contact process. In some cases, the contact can be eliminated by increasing the flow velocity. Throughout this section, we discuss only those cases with small $L^*=[0.6, 0.7]$ and $d^*=0.40$--0.55, which produce stronger sheet--wall interactions than other cases.

\citet{HyeonseongNano2020} categorised the wall contact mode of a three-dimensional sheet for $L^*=0.75$, $d^*=0.43$--0.60, and $U^*=9.4$--13.0 into three regimes of rolling, head-on, and touch/sliding contact, based on the position of the contact point on the sheet and the temporal variation in the contact force. Although these three regimes are also observed in our simulations, it is hard to determine the regime in some cases. Moreover, this categorisation is inappropriate for our two-dimensional sheet at relatively low Reynolds number and does not cover the wide ranges of parameters considered in our study. Instead, we propose four contact-mode regimes suitable for our model which is more general and easier to identify: sliding/rolling (type I), combination of sliding/rolling and bouncing (type II), bouncing (type III), and short touch (type IV), which embrace the regimes used by \citet{HyeonseongNano2020}. To identify the regime for each case, the morphing sequence of the sheet near contact and the time history of contact force are examined. The sheet is in contact with the wall if it has a non-zero contact force. A bounce is assumed to occur in two circumstances; first, if the sheet is contact with the wall for a time interval shorter than 0.5, and second, if the sheet looses contact (zero contact force) and contacts again with the same wall. Furthermore, if the sheet stays in contact with the wall for an interval greater than 0.5, the contact type is identified as rolling/sliding.

\begin{figure}
\centering
\includegraphics{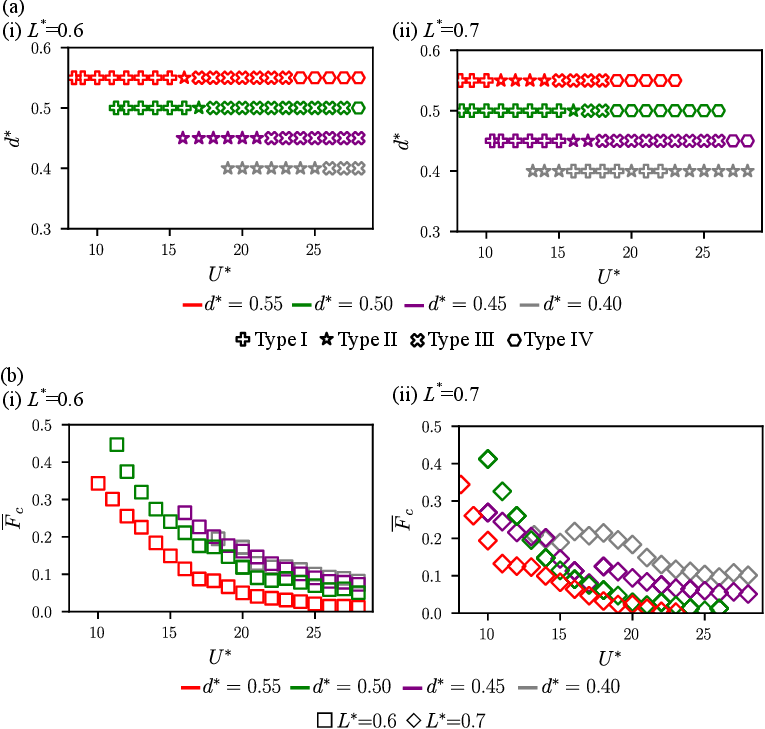}
\caption{(\emph{a}) Contact-mode distribution for (i) $L^*=0.6$ and (ii) $L^*=0.7$.  (b) Contact force coefficient $\overline{F}_c$ with respect to $U^*$ for (i) $L^* = 0.6$ and (ii) $L^* = 0.7$.}
\label{Contact_Regimes_Fc}
\end{figure}

\begin{figure}
\centering
\includegraphics{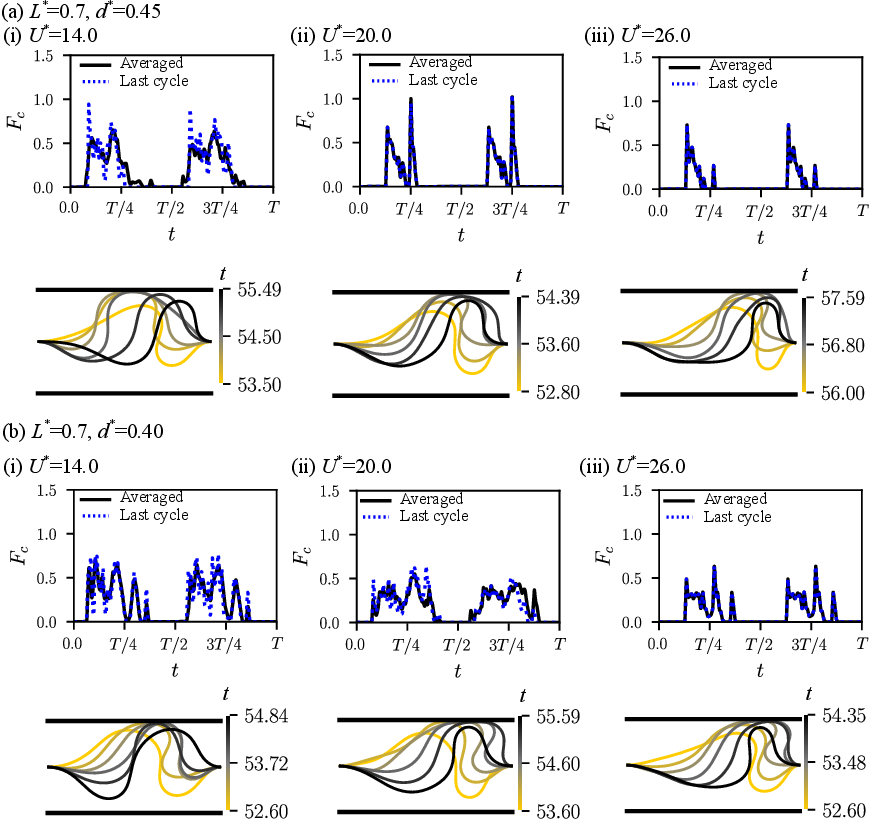}
\caption{Temporal variations in phase-averaged contact force $F_c$ for snap-through cycles between $t=20.0$--60.0 (black solid line) and instantaneous contact force in the last complete cycle before $t=60.0$ (blue dotted line), and sequences of sheet shape at $U^*=14.0$, 20.0, and 26.0: (\emph{a}) $L^* = 0.7$, $d^*=0.45$ and (\emph{b}) $L^* = 0.7$, $d^*=0.40$. For the shape morphing of each case, see supplementary movie 2.}
\label{contact_LdUf_Effects}
\end{figure}

Figure~\ref{Contact_Regimes_Fc}(\emph{a}) presents the distribution of the four contact modes for $L^*=[0.6, 0.7]$ and $d^*=0.40$--0.55. Near the critical velocity $U^*=U_c^*$, the oscillating sheet tends to exhibit a relatively long contact time ($\geq 0.5$) with the wall in the form of sliding or rolling (type I), which is similar to the behaviour in figure~\ref{contact_LdUf_Effects}(\emph{a}i) and supplementary movie 2(\emph{a}i). When $U^*$ increases, the sheet experiences a single or multiple bounces in addition to rolling/sliding motion, which is regarded as the combined mode (type II); this mode is observed for the upper-wall contact in figure~\ref{pressure_asymmetryShapes_d04}(\emph{b}) and supplementary movie 1(\emph{b}). A further increase in $U^*$ gradually weakens the sliding/rolling contact while increasing the number of bounces, and eventually the sheet only bounces multiple times near each of the confining walls (type III) (figure~\ref{contact_LdUf_Effects}\emph{a}iii and supplementary movie 2\emph{a}iii). For larger values of $U^*$, the number of bounces decreases until there is only a single short touch onto each confining wall (type IV), and complete elimination of the contact occurs for $d^*=0.50$ and 0.55, which is depicted as void in figure~\ref{Contact_Regimes_Fc}(\emph{a}ii).

The contact force coefficient $\overline{F}_c$ in figure~\ref{Contact_Regimes_Fc}(\emph{b}) is the contact force magnitude $F_c$ at both confining walls, which is time-averaged for all of the complete snap-through cycles within $t=20.0$--60.0. Note that we do not use the contact force averaged over a single cycle for the definition of $\overline{F}_c$. Averaging over a given time span is adopted, instead of averaging over a single cycle. It is because, in energy harvesting applications, it is important to produce larger contact forces and more frequent events of contact in a given time span. For given $L^*$ and $d^*$, a monotonic decrease in $\overline{F}_c$ with $U^*$ is generally observed. When $U^*$ is slightly greater than $U^*_c$,  sliding/rolling-based contact mode (type I or type II) occurs generally (figure~\ref{Contact_Regimes_Fc}\emph{a}). In this velocity regime, the sheet is in contact with the walls for a long time within the snap-through period (=$1/f^*$) and generates a large $\overline{F}_c$, indicating strong sheet--wall interaction. With increasing $U^*$, it becomes easier for the fluid flow to separate the sheet from the wall during the contact process, leading to the appearance of bounces. Therefore, the sheet--wall interaction weakens, and the total contact duration with non-zero contact force and the time-averaged contact force coefficient decrease.

\begin{table*}\caption{Contact force coefficient $\overline{F_c}$ and snap-through frequency $f^*$ of the dominant DMD mode for the cases in figure~\ref{contact_LdUf_Effects} } \label{tab:contact_LdUf_Effects}
\begin{center}
\begin{tabular}{ c c c c c c c}
$L^*$ & $d^*$ & $U^*$ & $\overline{F_c}$ & $f^*$ \\ [2ex]
0.7 & 0.45 & 14 & 0.18 & 0.125 \\
0.7 & 0.45 & 20 & 0.10 & 0.150 \\
0.7 & 0.45 & 26 & 0.06 & 0.153 \\[2ex]
0.7 & 0.40 & 14 & 0.19 & 0.112 \\
0.7 & 0.40 & 20 & 0.18 & 0.148 \\
0.7 & 0.40 & 26 & 0.10 & 0.150\\
\end{tabular}
\end{center}
\end{table*}

Figure~\ref{contact_LdUf_Effects} exemplifies the effects of increasing $U^*$ for $d^* = 0.45$ and $0.40$ ($L^*=0.7$). The time history of the contact force is phase-averaged for all complete cycles between  $t=20.0$--60.0 and plotted together with the contact force of the last cycle before $t=60.0$ in order to show the deviation of the instantaneous contact force from the phase-averaged value and the degree of repeatability. By increasing $U^*$, the deviation reduces and almost vanishes eventually, indicating that the contact force tends to have a complete periodic behaviour. This is accompanied by weaker contact force and generally shorter contact time relative to the snap-through period, and consequently smaller contact force coefficient.

In the examples of $L^* = 0.7$ and $d^* = 0.45$ (figure~\ref{contact_LdUf_Effects}(\emph{a}) and table~\ref{tab:contact_LdUf_Effects}), the wall exerts a force of $\overline{F}_c=0.18$ on the sheet at $U^* = 14.0$, corresponding to contact type I. Moreover, $\overline{F}_c$ falls to 56\% ($\overline{F}_c=0.10$) and then one-third of this value ($\overline{F}_c=0.06$) as $U^*$ increases to $U^*=20.0$ and $U^*=26.0$ with contact type III, respectively. On the other hand, all three cases in figure~\ref{contact_LdUf_Effects}(\emph{b}) have sliding/rolling-based contact modes (types I and II) and relatively long contact times with respect to their own snap-through period. Nevertheless, the temporal contact force changes notably as $U^*$ increases from 14.0 to 26.0 (figure~\ref{contact_LdUf_Effects}\emph{b}i--\emph{b}iii). When the time histories of the contact force are compared, it is evident that the contact force becomes weaker as the velocity increases from $U^*$ = 14.0 to 20.0 and the contact time within one snap-through period remains similar. However, $\overline{F}_c$ remains close for $U^* =$ 14.0 and 20.0 (table~\ref{tab:contact_LdUf_Effects}). This result is attributed to the significant increase in the snap-through frequency $f^*=f_d(2w_0)/U$ (defined in \S\ref{subsec:Dynamic mode decomposition}) from $0.112$ to $0.148$, which compensates for the reductions in peak contact force and contact time by increasing the number of contact events over a given time span; the snap-through frequency $f^*$ will be discussed in detail in \S\ref{subsubSec:Snap frequency}.

Reducing $L^*$ or $d^*$ (increasing $w_0^*$) at a given $U^*$ is expected to strengthen the sheet--wall interaction, producing a greater contact force coefficient $\overline{F}_c$. This is true for most of the cases examined in this study (figure~\ref{Contact_Regimes_Fc}\emph{b}). However, there are some exceptions due to the significant confinement of the confining walls. For example, for the smallest gap distance ratio $d^*=0.40$, when $L^*$ decreases from 0.7 to $0.6$, $\overline{F}_c$ remains similar or becomes somewhat smaller for most values of $U^*$. A decrease in $L^*$ to 0.6 causes an excessive blockage in the channel ($w_0^*=0.87$), which alleviates the fluid force exerted on the sheet and contributes to a decrease in the snap-through frequency $f^*$. Therefore, the decrease in $f^*$ is mainly responsible for the smaller $\overline{F}_c$.

As another example, for $L^*=0.7$, $\overline{F}_c$ has a negligible change from 0.18 to 0.19 at $U^* = 14.0$ as $d^*$ decrease from 0.45 to 0.40. Although the contact time is extended for $d^* = 0.40$ as illustrated in figure~\ref{contact_LdUf_Effects}(\emph{a}i,\emph{b}i), the smaller snap-through frequency and peak contact force of $d^* = 0.40$ cause the similarity in $\overline{F}_c$ between $d^*=0.45$ and $d^* = 0.40$. However, this trend at $U^*=14.0$ is not observed at larger values of $U^*$. At $U^*=20.0$ and 26.0, although the peak contact force is still lower for $d^*$ = 0.40 than for $d^*$ = 0.45, sliding/rolling-based contact modes (types I and II) with $d^*=0.40$ provide a longer contact time, contributing to the notably greater $\overline{F}_c$ (figure~\ref{contact_LdUf_Effects}\emph{b}ii,\emph{b}iii). By comparison, the bouncing contact mode (type III) with a reduced contact time is observed for $d^*$ = 0.45 at the same velocities $U^*=20.0$ and 26.0 (figure~\ref{contact_LdUf_Effects}\emph{a}ii,\emph{a}iii). In summary, neither $L^*$ nor $d^*$ has a simple monotonic relation with the contact force coefficient $\overline{F}_c$. Thus, the effects of $L^*$ and $d^*$ on the contact time, frequency, and peak contact force should be comprehensively considered to analyse the contact force coefficient.


\subsubsection{Snap-through frequency}
\label{subsubSec:Snap frequency}

\begin{figure}
\centering
\includegraphics{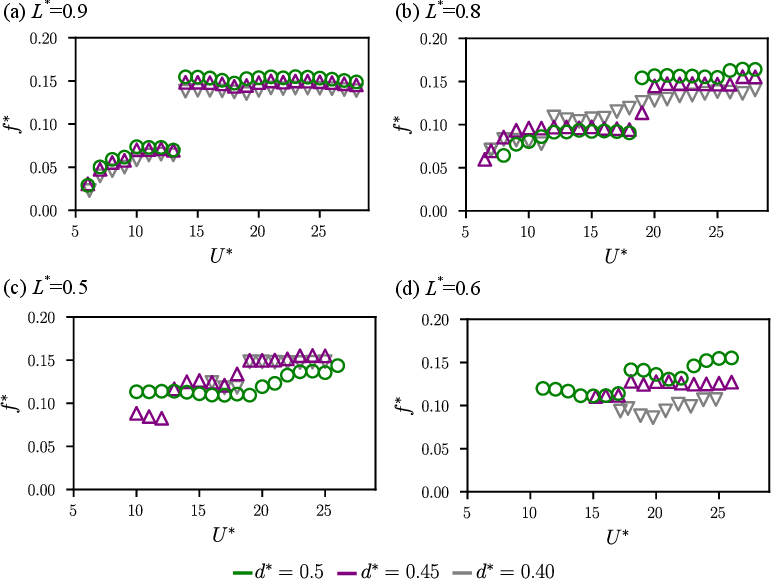}
\caption{Dimensionless snap-through frequency $f^*(=f_d(2w_0)/U)$ with respect to $U^*$ for different length ratios $L^*$.}
\label{fSnap}
\end{figure}

Some notable features of the frequency $f^*$ of snap-through oscillations can be identified by varying $L^*$, $d^*$, and $U^*$, as shown in figure~\ref{fSnap} for five cases of $d^*$, including the unbounded condition $d^* = \infty$. For given $L^*$ and $d^*$, the snap-through frequency generally increases with $U^*$ and is distributed in a narrow band at a high-$U^*$ regime of $U^*>22$. The frequencies for most values of $L^*$ and $d^*$ reside between 0.120--0.175 in this high-$U^*$ regime. For wall-bounded oscillations, \citet{HyeonseongNano2020} examined the snap-through frequency for a single length ratio of $L^*=0.75$ ($d^*=0.43$--0.60) in a limited range of $U^*=8.8$--13.0, and found a gradual increase in $f^*$ with respect to $U^*$. However, in our simulations considering wider ranges of $L^*$ and $U^*$, the case of $L^* = 0.9$ without contact exhibits two distinct frequency regimes divided by a sudden jump in the frequency at $U^* = 13$ (figure~\ref{fSnap}\emph{a}). For $L^* = 0.8$, the sudden jump in the frequency disappears as $d^*$ drops below $0.45$ and contact occurs; the frequency jump is also absent, and no consistent trend in $f^*$ can be observed, for $L^* = 0.7$ and 0.6. That is, large values of both $L^*$ and $d^*$ without contact are prone to sudden frequency jumps.

\begin{figure}
\centering
\includegraphics{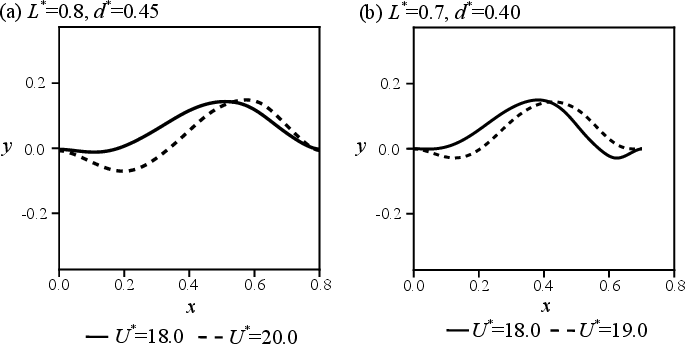}
\caption{Dominant DMD modes for (\emph{a}) $L^*=0.8$, $d^*=0.45$ and (\emph{b}) $L^*=0.7$ and $d^*=0.40$.}
\label{frequencyJump}
\end{figure}

Although not as steep as the cases in figure~\ref{fSnap}(\emph{a}), a notable frequency rise over a certain flow-velocity range also occurs in contact cases. To elaborate the cause of this frequency rise, the dominant DMD mode is illustrated in figure~\ref{frequencyJump}(\emph{a}) at $U^* = 18.0$ and 20.0 for $L^*=0.8$ and $d^*=0.45$. Only one dominant DMD mode is depicted because of its large amplitude compared with the other DMD modes. Even with a minor increase in $U^*$ from 18.0 to 20.0, the dominant mode shifts to one with a distinct nadir on the front part of the sheet and an apex displaced to the rear (figure~\ref{frequencyJump}\emph{a}). As explained in \S\ref{subSec:Equilibrium state and critical UStar}, the formation of a deep nadir on the front part annihilates the resistance of the sheet against the flow and precipitates snap-through. In the post-equilibrium state, the deep nadir in the dominant oscillatory mode contributes to the notable frequency rise from $f^* = 0.094$ to $0.145$. This drastic change in the dominant mode shape across a certain range of $U^*$ can occur for any $L^*$ and $d^*$, serving as a sufficient condition for the frequency rise. This phenomenon is also responsible for the sudden frequency jump shown in figure~\ref{fSnap}(\emph{a}).

Another type of mode shape change leading to a notable frequency rise is shown for $L^*=0.7$ and $d^*=0.40$ (figure~\ref{frequencyJump}\emph{b}). The dominant mode at $U^*=19.0$ has a nadir on the rear part of the sheet, which is not present in figure~\ref{frequencyJump}(\emph{a}). This dominant mode appears for all cases with $w_0^*\geq0.66$. In \S\ref{subSec:Equilibrium state and critical UStar}, it was reported that the shape of the sheet in the equilibrium state has a nadir on the rear part for $w_0^*\geq0.66$ and a nadir on the front part for $w_{0}^*<0.66$ (table~\ref{tab:shapeshift_dStar}). This indicates that the quasi-static shape of the sheet in the equilibrium state is closely related to the dynamic mode in the post-equilibrium state for given $L^*$ and $d^*$. However, the dominant mode with a nadir on the rear part disappears with a further increase in $U^*$. Indeed, at a slightly greater flow velocity of $U^*$ = 20 (figure~\ref{frequencyJump}\emph{b}), the dominant mode shifts to a shape with a nadir on the front part, resulting in a notable frequency rise from $f^* = 0.120$ to 0.152.

\subsection{Flow structure in post-equilibrium state}
\label{subSec:Flow structure}

In this section, for the first time, the flow structure around and behind the oscillation sheet is analyzed in detail for the cases of unbounded and wall-bounded snap-through. Various gap distances between the confining walls, the smallest length ratio of $L^*=0.6$ and a dimensionless velocity of $U^*=24.0$ are chosen to address the effects of the confining walls in comparison with the unbounded case. These cases also exemplify the common features of the flow structure, including the mechanism of vortex formation, dissipation in the presence of the confining walls, and the effects of the sheet motion. In our model at a low Reynolds number ($Re = 200$) with strong viscous diffusion, vortices develop separately around four points of the sheet without the confining walls: two clamped ends of the sheet and two extrema (apex and nadir) formed by the oscillation of the sheet in the upper and lower regions of the channel.

\begin{figure}
\centering
\includegraphics{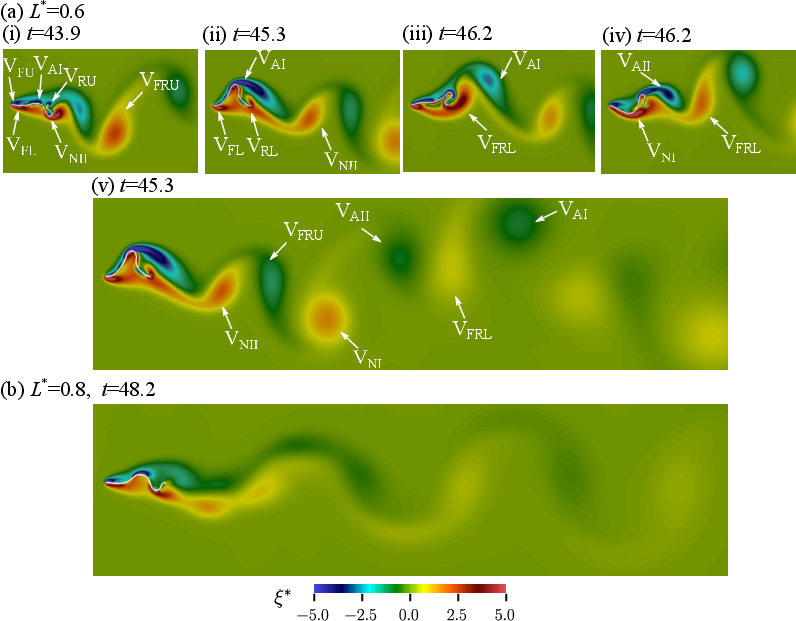}
\caption{Sequential snapshots of normalised vorticity field $\xi^*=\xi w_0/U$ for $d^*=\infty$ (unbounded) with (\emph{a}) $L^*=0.6$ and (\emph{b}) $L^*=0.8$ ($U^* = 24.0$). The colour bar corresponds to the values of $\xi^*$. See supplementary movie 3.}
\label{vorticityUnbounded}
\end{figure}

Figure~\ref{vorticityUnbounded}(\emph{a}i) shows the instant when the unbounded sheet with $d^* = \infty$ is moving upwards and has just passed the centreline; supplementary movie 3. Two counter-rotating vortical regions with similar magnitudes form around the front end of the sheet on the upper (V\textsubscript{FU}) and lower (V\textsubscript{FL}) sides. These two vortices stretch along with the sheet and supply vorticity to the vortices generated from the apex, nadir, and rear end of the sheet. A positive vortex that has already formed below the nadir, V\textsubscript{NII}, begins to detach from the sheet. As the apex is forming on the upper side (figure~\ref{vorticityUnbounded}\emph{a}ii), a negative vortex, V\textsubscript{AI}, emerges above the apex. Next, when the apex shifts along the streamwise direction on the upper side (figure~\ref{vorticityUnbounded}\emph{a}iii), the vortices generated from the front and rear ends on the lower side (V\textsubscript{FL}, V\textsubscript{RU}) merge to form one positive vortex, V\textsubscript{FRL}. The first apex vortex, V\textsubscript{AI}, interacts with the counter-rotating V\textsubscript{FRL} and sheds from the apex. Simultaneously, a second apex vortex, V\textsubscript{AII}, forms above the apex. Along with the snapping motion of the sheet to the lower side (figure~\ref{vorticityUnbounded}\emph{a}iv), V\textsubscript{AII} stretches downstream and induces the detachment of V\textsubscript{FRL}. V\textsubscript{AII} also begins to detach after interacting with a positive vortex, V\textsubscript{NI}, that forms below the developing nadir on the lower side.

\begin{figure}
    \centering
    \includegraphics{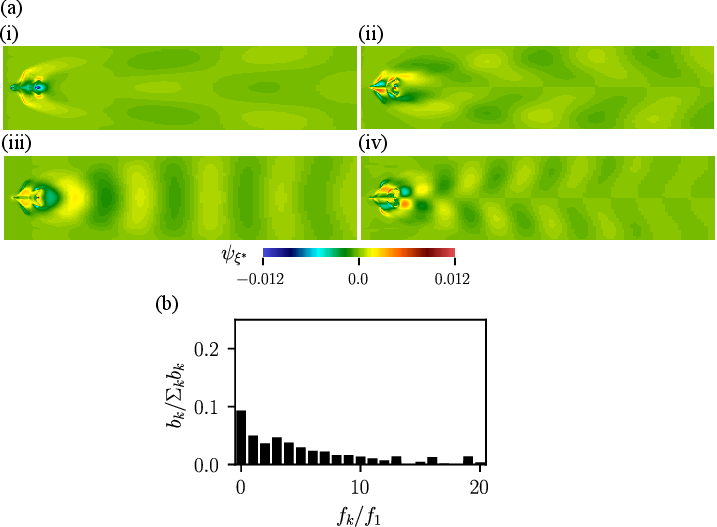}
    \caption{(\emph{a}) DMD modes of the flow field for the unbounded case ($L^*=0.6$, $U^* = 24.0$): (i) first (fundamental), (ii) second, (iii) third, and (iv) fourth modes. (\emph{b}) Coefficients of the DMD modes normalised by the sum of all modes. The colourbar corresponds to the values of the DMD modes for the vorticity field, $\psi_{\xi^*}$ in (\emph{a}).}
    \label{DMD_00}
\end{figure}

The dynamics of the sheet, which drive the formation and detachment of the vortices, are critical in determining the distribution of flow structures in the wake. This is also supported by the DMD analyses of the sheet shape and vorticity field in the fluid domain of $[-0.2,8.0]\times [-1.0,1.0]$. The morphing of the sheet excites several harmonics of the vorticity field in the wake, and the fundamental frequency of the vorticity field is equal to that of the sheet motion. Figure~\ref{DMD_00}(\emph{a}) shows the first four DMD modes of the vorticity field. The first DMD mode is symmetric as in the first mode of vortex shedding around a bluff body (e.g. circular cylinder) \citep{Bagheri2013}. The value of the DMD mode is largest near the two clamping ends. The second DMD mode is antisymmetric, vorticities of opposite signs, which are formed in the middle of the sheet on both upper and lower sides of the centreline, are clearly visible, highlighting strong coupling between the dynamics of the sheet and flow structure. Although the amplitudes (coefficients $b_k$ in~\eqref{DMDSolution}) of the higher frequencies generally tend to decrease, the amplitude of the third mode, which contains the symmetric footprint of vortices, is comparable to that of the first mode, and the amplitude of the fourth mode, having the antisymmetric distribution, is also comparable to the second mode (figure~\ref{DMD_00}\emph{b}).

In each cycle of unbounded snap-through oscillations with $L^*=0.6$, six distinct vortices are shed in the wake: two from the apex (V\textsubscript{AI}, V\textsubscript{AII}), two from the nadir (V\textsubscript{NI}, V\textsubscript{NII}), and two produced by the front and rear vortices merging on the upper and lower sides, respectively (V\textsubscript{FRU}, V\textsubscript{FRL}) (figure~\ref{vorticityUnbounded}\emph{a}v; note that only a half period is depicted in figure~\ref{vorticityUnbounded}\emph{a}i--\emph{a}iv). This flow pattern is utterly different from those of the other common configurations of the sheet, i.e. fluttering flags in which dominant vortices form and detach from the free leading or trailing edge~\citep{Shoele2016,Kim2013}. Although the vortices develop around the extrema of the sheet for all values of $L^*$, they become weaker and smaller for larger $L^*$ because of the reduced transverse displacement of the sheet (figure~\ref{vorticityUnbounded}\emph{b}). Detached weak vortices dissipate quickly, and their footprint is visible in the far wake region only as stretched vortical regions with small vorticity magnitudes.

\begin{figure}
\centering
\includegraphics{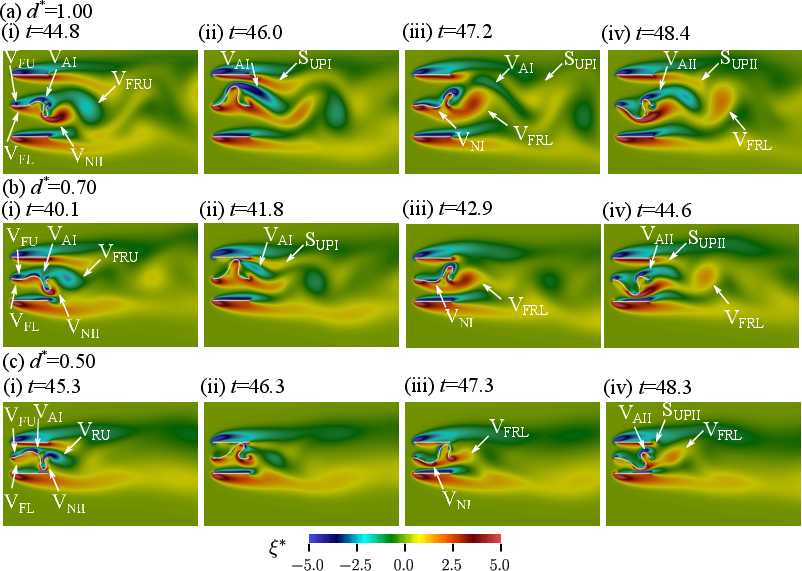}
\caption{Sequential snapshots of normalised vorticity field $\xi^*=\xi w_0/U$ near the sheet for (\emph{a}) $d^*=1.00$, (\emph{b}) $d^*=0.70$, and (\emph{c}) $d^*=0.50$ ($L^* = 0.6$, $U^* = 24.0$). The colour bar corresponds to the values of $\xi^*$. See supplementary movie 4.}
\label{vorticityCloseViewWallBounded}
\end{figure}

The placement of the confining walls does not apparently affect the general pattern of vortex formation around the sheet. For example, in figure~\ref{vorticityCloseViewWallBounded}(\emph{a}i,\emph{b}i,\emph{c}i) when the sheet is moving upwards and the apex is forming, the front vortices V\textsubscript{FU}/V\textsubscript{FL}, the first apex vortex V\textsubscript{AI}, and the second nadir vortex V\textsubscript{NII}, as well as the negative rear vortex V\textsubscript{RU}, are all present, similar to the unbounded case in figure~\ref{vorticityUnbounded}(\emph{a}i); supplementary movie 4. This observation is somehow expected because the vortex formation around the sheet is driven by the morphological change of the sheet over time. However, the placement of the confining walls creates two shear layers on the top and bottom surfaces of each confining wall. The shear layers on the inner surfaces of the confining walls interact with the flow around the sheet and alter the growth and shedding of the vortices from the sheet.

When the confining walls are positioned far from the sheet without contact (i.e. $d^*$ = 1.00), the first apex vortex, V\textsubscript{AI}, forms above the apex, similar to the unbounded case, and interacts with the positive shear layer, S\textsubscript{UPI}, that emerges on the lower surface of the upper wall. In the absence of the sheet, S\textsubscript{UPI} stretches downstream to a distance of about $9L^*$ and remains stable. In contrast, being affected by the flow over the moving sheet, S\textsubscript{UPI} is compressed by V\textsubscript{AI} inside the channel, and is dragged downwards in the near wake (figure~\ref{vorticityCloseViewWallBounded}\emph{a}ii). Subsequently, S\textsubscript{UPI} detaches from the confining wall along with V\textsubscript{AI}, forming two counter-rotating stretched and thin vortices in the wake (figure~\ref{vorticityCloseViewWallBounded}\emph{a}iii). Closer placement of the confining walls ($d^*=0.70$) leads to stronger interaction between V\textsubscript{AI} and S\textsubscript{UPI} even without contact, which hinders the development and stretching of both vortices. As a result, compared with the larger gap of $d^* = 1.00$, V\textsubscript{AI} and S\textsubscript{UPI} become weaker near the channel, and dissipate immediately downstream (figure~\ref{vorticityCloseViewWallBounded}\emph{b}ii,\emph{b}iii). Furthermore, in the contact case of $d^*=0.50$, V\textsubscript{AI} lacks enough space to grow in the narrow gap between the upper wall and the sheet, and V\textsubscript{AI} and S\textsubscript{UPI} vanish following the impingement of the sheet onto the confining wall, leaving no clear wake structure (figure~\ref{vorticityCloseViewWallBounded}\emph{c}ii,\emph{c}iii).

When the sheet moves in the opposite direction towards the lower confining wall, another positive shear layer below the upper wall, S\textsubscript{UPII}, and a second apex vortex, V\textsubscript{AII}, begin to grow and interact with each other. By virtue of the downward motion of the sheet, V\textsubscript{AII} has enough space to form in all three representative cases. For $d^*=1.00$, V\textsubscript{AII} stretches downstream along with S\textsubscript{UPII} (figure~\ref{vorticityCloseViewWallBounded}\emph{a}iv), followed by their detachment from the sheet and the upper wall, respectively. Comparatively, in the cases of $d^*=0.70$ and 0.50, both S\textsubscript{UPII} and V\textsubscript{AII} exhibit less stretching and quickly become weaker (figure~\ref{vorticityCloseViewWallBounded}\emph{b}iv,\emph{c}iv). This behaviour stems from the delay in their initial growth due to the small gap between the sheet and the upper wall and the contact process. While the vortices formed from the apex of the sheet can be disrupted and their shedding is suppressed by narrowing the gap distance, the vortices originating at the front and rear ends merge to form V\textsubscript{FRL} below the sheet, with a sufficient distance from the lower wall. V\textsubscript{FRL} is convected downstream, preserving a clearly identifiable core (figure~\ref{vorticityCloseViewWallBounded}\emph{a}iv,\emph{b}iv,\emph{c}iv).

\begin{figure}
\centering
\includegraphics{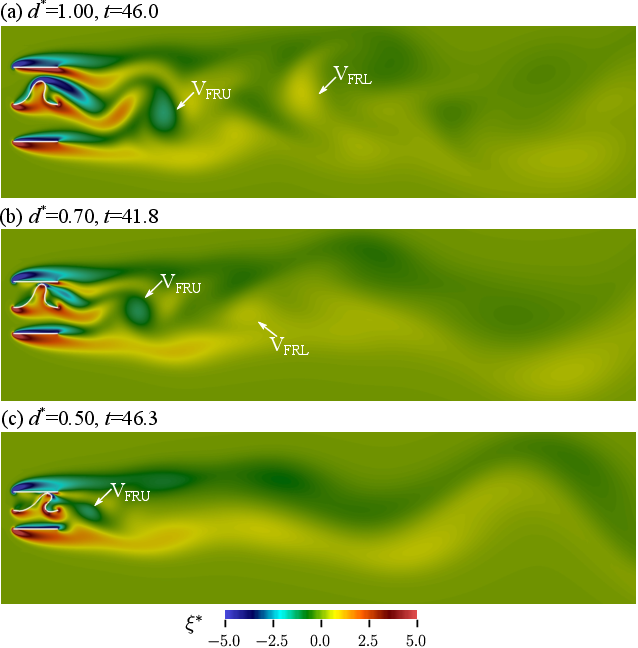}
\caption{Snapshots of normalised vorticity field $\xi^*=\xi w_0/U$ in the far wake for (\emph{a}) $d^*=1.00$, (\emph{b}) $d^*=0.70$, and (\emph{c}) $d^*=0.50$ ($L^* = 0.6$, $U^* = 24.0$). The colour bar corresponds to the values of $\xi^*$. }
\label{vorticityWakeWallBounded}
\end{figure}

For $d^*=1.00$, six vortices are shed from the sheet in each cycle, similar to the unbounded case. Among them, V\textsubscript{FRL} and V\textsubscript{FRU} are the strongest and are convected farther downstream without losing their forms (figure~\ref{vorticityWakeWallBounded}\emph{a}). However, as $d^*$ decreases to 0.70 and 0.50, the vortices that separate from the sheet are weak and dissipate more quickly. Accordingly, in the far-wake region, the shear layers from the outer surfaces of the confining walls are prevalent over the vortices generated by the sheet (figure~\ref{vorticityWakeWallBounded}\emph{b},\emph{c}). Because of the interaction with the vortices shed from the sheet in the near-wake region, the shear layers become unstable, undulating in the far-wake region; note that this instability of the shear layers does not occur in the absence of the sheet. These unstable shear layers appear for all values of $L^*$ and small values of $d^*$.

\begin{figure}
    \centering
        \includegraphics{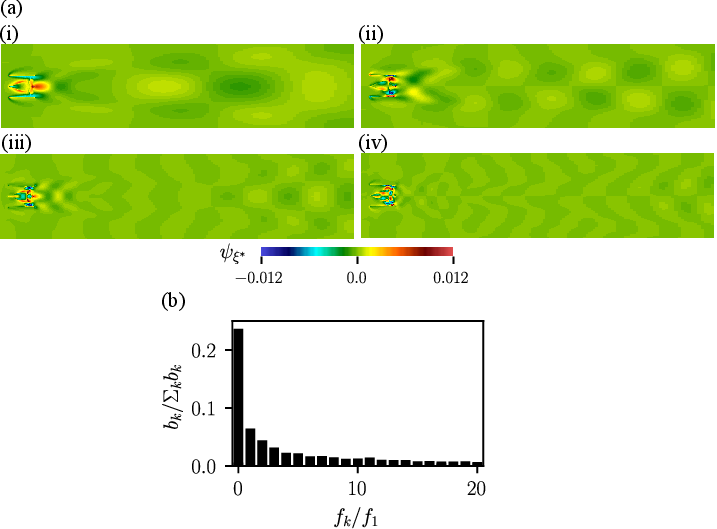}
    \caption{(\emph{a}) DMD modes of the flow field for the wall-bounded case ($L^*=0.6$, $d^*=0.5$, $U^*=24.0$): (i) first (fundamental), (ii) second, (iii) third, and (iv) fourth modes. (\emph{b}) Coefficients of the DMD modes normalised by the sum of all modes. The colourbar corresponds to the values of the DMD modes for the vorticity field, $\psi_{\xi^*}$ in (\emph{a}).}
    \label{DMD_05}
\end{figure}

Similar to the unbounded case, the fundamental frequency of the vorticity field from the DMD analysis is identical to that of the sheet motion, although the wake structure of the contact cases is dominated by the unstable shear layers from the confining walls rather than the vortices that are periodically shed from the sheet. This finding is supported by the DMD modes of the vorticity field (figure~\ref{DMD_05}\emph{a}). Vorticities generated on the walls are pronounced in the first DMD mode, and vorticities of opposite signs formed in the middle of the sheet are prevalent in the second DMD mode. Particularly, in the second to fourth modes, the region of strong vorticity on the inner surface of the confining wall indicates its role in interrupting the development of the nearby counter-rotating vortex formed on the sheet. Because of the diffusive effect by the presence of the walls, less interaction occurs between the sheet vortices. Accordingly, the amplitudes of the higher modes drop quickly after the first mode (figure~\ref{DMD_05}\emph{b}). In this case, the amplitudes of the third and fourth modes are notably smaller than those of the first and second modes, respectively; note the distinct difference in the amplitude distribution for the DMD modes between figures~\ref{DMD_00}(\emph{b}) and~\ref{DMD_05}(\emph{b}).

\section{Concluding remarks}
\label{sec:Concluding remarks}

We have numerically investigated the snap-through oscillations of a two-dimensional sheet confined between two confining walls, revealing novel features of the sheet motion and flow structure, which have not been reported before. The equilibrium shape of the sheet in contact with the confining walls deviates from that of non-contact cases. The nadir formed on the front part of the sheet by the contact with the confining wall becomes more pronounced with decreasing gap distance, and causes a reduction in the critical flow velocity $U^*_c$. However, below a critical gap distance which is correlated with the blockage ratio, the nadir shifts to the rear part of the sheet, and the distribution of the net pressure force applied to the sheet changes remarkably, leading to a rise in $U^*_c$. The post-equilibrium state of contact cases generally begins with rolling/sliding-based contact at flow velocities close to $U^*_c$, which features a longer contact time and a greater contact force coefficient compared with the bouncing-type contact cases at larger values of $U^*$. Furthermore, the contact force generally strengthens with smaller length ratios and gap distance ratios, although some exceptional cases exist at extreme blockage ratios. For all cases of $L^*$ and $d^*$, the sudden rise in the snap-through frequency with increasing $U^*$ is accompanied by a shift in the dominant oscillatory mode to a shape with a nadir formed on the left and an apex displaced in the streamwise direction. Bringing the confining walls closer together disrupts the periodic vortex shedding from the oscillating sheet and causes the wake structure to be dominated by the shear layers created by the confining walls, rather than the vortices created by the sheet. Despite strong dissipation of the vortices by the confining walls, the fundamental frequency in the dynamic mode of the wake structure is determined by the snap-through frequency of the sheet.

Although this study has been limited to two-dimensional and laminar flow assumptions, which are far from the conditions of actual energy harvesting applications, it has covered several important aspects of the periodic snap-through that can be exploited to improve the design and performance of triboelectric energy generation. The snap-through oscillations of a buckled sheet under interactions with nearby objects induce interesting phenomena which deepen our knowledge in regards to flow-induced vibrations. Future studies of snap-through oscillations need to consider the effects of different inclination angles at two clamped ends of the sheet and the mutual interaction of multiple sheets in side-by-side arrangements.

\section*{Acknowledgements}
This research was supported by the Basic Science Research Program through the National Research Foundation of Korea (NRF) funded by the Ministry of Science and ICT (NRF-2020R1A2C2102232).

\section*{Declaration of interests}
The authors report no conflict of interest.

\bibliographystyle{jfm}
\bibliography{Snap_through}

\end{document}